\documentclass{article}
\usepackage{geometry}
\usepackage{graphics}

\makeatletter

\providecommand{\LyX}{L\kern-.1667em\lower.25em\hbox{Y}\kern-.125emX\@}

\makeatother

\begin{document}

\title{Coherent backscattering of light in a magnetic field}

\author{D. Lacoste and B. A. van Tiggelen \\
  \emph{Laboratoire de Physique et Mod\'{e}lisation des Milieux Condens\'{e}s,
}\\
\emph{CNRS, Université Joseph Fourier}, \emph{Maison des Magist\`{e}res, }\\
 \emph{}B.P. 166, 38042 Grenoble Cedex 09, France \\
 }

\maketitle
\begin{abstract}
This paper describes how coherent backscattering is altered by an external magnetic
field. In the theory presented, magneto-optical effects occur inside Mie scatterers
embedded in a non-magnetic medium. Unlike previous theories based on point-like
scatterers, the decrease of coherent backscattering is obtained in leading order
of the magnetic field using rigorous Mie theory. This decrease is strongly enhanced
in the proximity of resonances, which cause the path length of the wave inside
a scatterer to be increased. Also presented is a novel analysis of the shape
of the backscattering cone in a magnetic field. 
\end{abstract}

\section{Introduction}

The enhancement of backscattering in propagation of waves in a random medium
is a well documented topic. Weak localization theory explains how interference
effects between direct and reverse scattering events produce Coherent Backscattering
(CBS). The main features of CBS are insensitive to many aspects of the statistics
of the inhomogeneities. Even absorption does not alter the relative strength
of the interference \cite{akkermans}. The interference is only affected by
the properties of the medium with respect to the reciprocity principle as first
noted by Golubentsev \cite{golubentsev}. MacKintosh and John \cite{john} analyzed
how CBS is altered by Faraday rotation and natural optical activity in a medium
of inhomogeneities smaller than the wavelength, and in the diffusion approximation.
Using a method based on point-like scatterers, Van Tiggelen \emph{et al.} extended
these ideas and discussed the anisotropy induced by a magnetic field in light
diffusion \cite{bart}. The case of non-magnetic Mie scatterers embedded in
a Faraday-active medium has been studied by means of Monte Carlo simulations
by Martinez \emph{et al.} \cite{martinez}. Maradudin \emph{et al.} considered
specifically the two dimensional coherent backscattering of light from a randomly
rough surface in the presence of a magnetic field \cite{maradudin}. Experimentally,
CBS in a magnetic field has been studied by Erbacher, Lenke and Maret \cite{lenke},
and some of their results will be discussed here. 

In section \ref{Sec:T_matrix}, the main results of a recent calculation of
the T-matrix for a Mie scatterer in a magnetic field are presented \cite{josa,jqsr},
and serve in section \ref{Sec:diffusion} as the building block to study diffusion
of light in a magnetic field. After having detailed the main features of the
Faraday effect for multiple Rayleigh and Mie scattering, the modification of
the line shape of CBS in a medium with finite-size scatterers in a magnetic
field is investigated in the last section.

\section{T-matrix in a magnetic field}

\label{Sec:T_matrix}In this paper, \( c_{0}=1 \) has been set. In a magnetic
field, the refractive index is a tensor of rank two. For the standard Mie problem,
it depends on the distance to the center of the sphere \( r \), which has a
radius \( a \), via the Heaviside function \( \Theta (r-a), \) that equals
1 inside the sphere and 0 outside, \textbf{
\begin{eqnarray}
\varepsilon (\mathbf{B},\mathbf{r})-\mathbf{I}= & \left[ (\varepsilon _{0}-1)\, 
\mathbf{I}+\varepsilon _{F}\, \, \Phi \right] \Theta (r-a).\label{epsilon} 
\end{eqnarray}
}In this expression, \( \varepsilon _{0}=m^{2} \) is the value of the normal
isotropic dielectric constant of the sphere of relative index of refraction
\( m \), and 
\begin{equation}
\label{epsF}
\varepsilon _{F}=2mV_{0}B/\omega 
\end{equation}
 is a dimensionless coupling parameter associated with the amplitude of the
Faraday effect (\( V_{0} \) being the Verdet constant, \( B \) the amplitude
of the magnetic field and \( \omega  \) the frequency). The antisymmetric hermitian
tensor \( \Phi _{ij}=i\epsilon _{ijk}\hat{B}_{k} \) has been introduced (the
hat above vectors means that the vectors have been normalized). The Mie solution
depends on the dimensionless size parameters \( x=\omega a \) and \( y=mx \).
In this paper, only non-absorbing media are considered so that \( m \) and
\( \varepsilon _{F} \) are real-valued. Since \( \varepsilon _{F}\approx 10^{-4} \)
in most experiments, a perturbational approach is valid. The part of \( \mathbf{T} \)
that is independent of the magnetic field is denoted \( \mathbf{T}^{0} \),
the part of the T-matrix linear in \( \mathbf{B} \) is \( \mathbf{T}^{1} \),
and the second order correction \( \mathbf{T}^{2} \).

Two important symmetry relations must be obeyed by a T-matrix of a scatterer
in a magnetic field. The first one is parity symmetry and the second one is
reciprocity \cite{josa}:

\begin{equation}
\label{parity}
T_{-\mathbf{k}\sigma ,-\mathbf{k}'\sigma '}(\mathbf{B})=T_{\mathbf{k}\sigma
 ,\mathbf{k}'\sigma '}(\mathbf{B}),
\end{equation}

\begin{equation}
\label{reciprocity}
T_{-\mathbf{k}'-\sigma ',-\mathbf{k}-\sigma }(-\mathbf{B})=T_{\mathbf{k}\sigma 
,\mathbf{k}'\sigma '}(\mathbf{B}).
\end{equation}
 It is important to note that \( \sigma (-\hat{\mathbf{k}})=-\sigma (\hat{\mathbf{k}}) \),
i.e. \( \sigma  \) indicates the helicity. In particular, relations (\ref{parity},
\ref{reciprocity})
must hold for \( \mathbf{T}^{1} \). \( \mathbf{T}^{2} \) satisfies Eq. (\ref{reciprocity})
without a minus sign for \( \mathbf{B} \) and obeys the standard reciprocity
principle. For this reason, \( \mathbf{T}^{2} \) will not contribute to the
suppression of the backscattering cone considered in this article, the next
order being \( \mathbf{T}^{3} \).

Because \textbf{\( \mathbf{T}^{1} \)} is linear in \( \hat{\mathbf{B}} \),
it can be constructed by considering only three special cases for the direction
of \( \hat{\mathbf{B}} \). If \( \hat{\mathbf{k}} \) and \( \hat{\mathbf{k}}' \)
are not collinear, the unit vector \( \hat{\mathbf{B}} \) can be decomposed
in the non-orthogonal but complete basis of \( \hat{\mathbf{k}},\, \hat{\mathbf{k}}' \)
and \( \hat{\mathbf{g}}=\hat{\mathbf{k}}\times \hat{\mathbf{k}}'/|\hat{\mathbf{k}}
\times \hat{\mathbf{k}}'| \).
This results in \cite{josa},
\begin{eqnarray}
\mathbf{T}_{\mathbf{kk}'}^{1} & = & \frac{(\hat{\mathbf{B}}\cdot \hat{\mathbf{k}})
(\hat{\mathbf{k}}\cdot \hat{\mathbf{k}}')-\hat{\mathbf{B}}\cdot \hat{\mathbf{k}}'}{(\hat
{\mathbf{k}}\cdot \hat{\mathbf{k}}')^{2}-1}\mathbf{T}_{\mathbf{kk}'}^{1}(\hat{\mathbf{B}}
=\hat{\mathbf{k}}')+\frac{(\hat{\mathbf{B}}\cdot \hat{\mathbf{k}}')(\hat{\mathbf{k}}\cdot 
\hat{\mathbf{k}}')-\hat{\mathbf{B}}\cdot \hat{\mathbf{k}}}{(\hat{\mathbf{k}}\cdot \hat{
\mathbf{k}}')^{2}-1}\mathbf{T}_{\mathbf{kk}'}^{1}(\hat{\mathbf{B}}=\hat{\mathbf{k}})
\nonumber \\
 & + & (\hat{\mathbf{B}}\cdot \hat{\mathbf{g}})\mathbf{T}_{\mathbf{kk}'}^{1}
 (\hat{\mathbf{B}}=\hat{\mathbf{g}}),\label{Tkkp} 
\end{eqnarray}

With respect to an helicity base, \( \mathbf{T}^{1} \) takes the form,

\begin{equation}
\label{tbpk}
T_{\sigma \sigma '}^{1}(\hat{\mathbf{B}}=\hat{\mathbf{k}})=\frac{\pi }{\omega }
[R_{1}(\theta )\sigma +R_{2}(\theta )\sigma '],
\end{equation}

\begin{equation}
\label{tbpkp}
T_{\sigma \sigma '}^{1}(\hat{\mathbf{B}}=\hat{\mathbf{k}}')=\frac{\pi }{\omega }
[R_{1}(\theta )\sigma '+R_{2}(\theta )\sigma ],
\end{equation}

 with

\begin{equation}
\label{p1}
R_{1}(\theta )=-\frac{2\varepsilon _{F}}{m}\sum _{J\geq 1}\frac{2J+1}{J(J+1)}\, 
[\mathcal{C}_{J}\pi _{J,1}(\theta )+\mathcal{D}_{J}\tau _{J,1}(\theta )]
\end{equation}

\begin{equation}
\label{p2}
R_{2}(\theta )=-\frac{2\varepsilon _{F}}{m}\sum _{J\geq 1}\frac{2J+1}{J(J+1)}\, 
[D_{J}\pi _{J,1}(\theta )+C_{J}\tau _{J,1}(\theta )]
\end{equation}
 where the coefficients \( \mathcal{C}_{J} \) and \( \mathcal{D}_{J} \) have
been defined in Ref. \cite{josa}. For \( \mathbf{T}_{\sigma \sigma '}^{1}
(\hat{\mathbf{B}}=\hat{\mathbf{g}}) \),
the following expression has been found 

\begin{equation}
\label{tbg}
T_{\sigma \sigma '}^{1}(\hat{\mathbf{B}}=\hat{\mathbf{g}})=\frac{\pi }{\omega }
(Q_{1}+\sigma \sigma 'Q_{2})
\end{equation}
 with

\begin{equation}
\label{q1}
Q_{1}=-i\frac{d}{d\theta }R_{1}\, ,\, \, \, \, \, \, \, \, \, \, \, \, \, \, \, 
\, \, Q_{2}=-i\frac{d}{d\theta }R_{2}.
\end{equation}
 For \( \hat{\mathbf{k}}=\hat{\mathbf{k}}' \),
\begin{equation}
\label{tkk}
\mathbf{T}_{\mathbf{k},\mathbf{k}}^{1}=\Phi \, \frac{2\pi }{\omega }R_{1}(0),
\end{equation}
 with

\begin{equation}
\label{p10}
R_{1}(0)=R_{2}(0)=-\frac{\varepsilon _{F}}{m}\sum _{J\geq 1}(2J+1)\, 
(\mathcal{C}_{J}+\mathcal{D}_{J}).
\end{equation}

\section{Diffusion of light in a magnetic field }

\label{Sec:diffusion}
\begin{figure}
{\par\centering \resizebox*{15cm}{15cm}{\rotatebox{-90}{\includegraphics{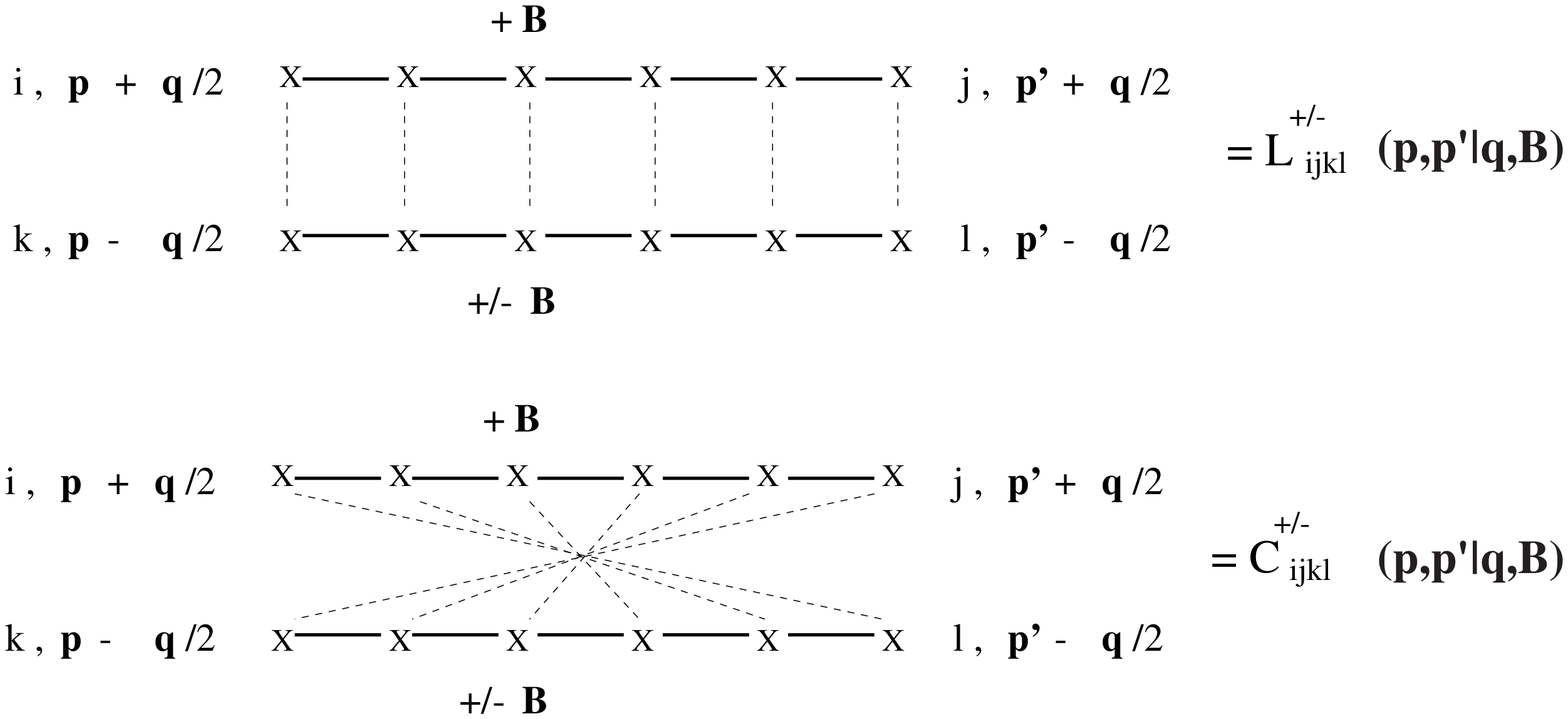}}} \par}

\caption{Ladder diagrams \protect\( L^{\pm }_{ijkl}\protect \) and the most-crossed
diagrams \protect\( C^{\pm }_{ijlk}\protect \) in a magnetic field. Bold lines
denote the Dyson-Green tensor, the crosses denote T-matrices and dotted lines
connect identical particles. \label{Fig:ladder}}
\end{figure}

The equation of radiative transfer aims to describe the
propagation of the average intensity in multiple scattering, but violates reciprocity
\cite{maynard}. The reason is that it does not contain the most-crossed diagrams
(\( C^{+}_{ijkl} \)) responsible for CBS. By definition, the four-rank tensor
\( L_{ijkl} \) linearly connects the field correlations of incident and outgoing
fields. This can be represented by a diagram that starts on the first scattering
event, and which has two lines, corresponding to the propagation of the field
and to the complex conjugate of the field as can be seen in Fig. (\ref{Fig:ladder}).
For \( L^{+}_{ijkl} \) both fields propagate in the same magnetic field \( \mathbf{B} \)
as opposed to \( L^{-}_{ijkl} \) for which both fields propagate with opposite
sign for the magnetic field \( \mathbf{B} \) \cite{bart}. The relations deduced
from the application of the reciprocity relation (\ref{reciprocity}) are

\begin{equation}
\label{CL}
C^{\pm }_{ijlk}(\mathbf{p},\mathbf{p}',\mathbf{q},\mathbf{B})=L^{\mp }_{ijkl}\left( 
\frac{\mathbf{p}-\mathbf{p}'+\mathbf{q}}{2},\frac{\mathbf{p}'-\mathbf{p}+\mathbf{q}}{2},
\mathbf{p}+\mathbf{p}',\mathbf{B}\right) ,
\end{equation}
and 

\begin{equation}
\label{LL}
L^{\pm }_{ijkl}(\mathbf{p},\mathbf{p}',\mathbf{q},\mathbf{B})=L^{\pm }_{jilk}\left( 
-\mathbf{p}',-\mathbf{p},-\mathbf{q},-\mathbf{B}\right) .
\end{equation}
 On long length scales (\( \mathbf{q}\rightarrow 0 \)) and for stationary situations
(\( \Omega =0 \)), the ladder diagrams \( L^{-}_{ijkl} \) take the following
hydrodynamic form \cite{euro},

\begin{equation}
\label{C}
L^{-}_{ijkl}(\mathbf{p},\mathbf{p}',\mathbf{q},\mathbf{B})=\frac{2\pi }{\ell ^{2}}
\frac{d_{ik}(\mathbf{p},\mathbf{q},\mathbf{B})d_{lj}(-\mathbf{p}',-\mathbf{q},
-\mathbf{B})}{D\mathbf{q}^{2}+\lambda /\ell }.
\end{equation}
 The symmetry of the numerator is imposed by the reciprocity principle; \( \lambda  \)
is a scalar dimensionless parameter that elucidates the breaking of reciprocity
by the magnetic field in multiple scattering. In Eq. (\ref{C}), \( D \) is
the conventional diffusion coefficient for radiative transfer, and \( \ell  \)
is the scattering mean free path, which is a typical distance between two scattering
events. This diffusion constant was shown theoretically and experimentally \cite{anja}
to depend on the square of the magnetic field. These corrections are not considered
here since the dephasing parameter \( \lambda /(D\ell ) \) is only discussed
to \emph{second} order in the magnetic field. The enhancement factor of coherent
backscattering is a ratio of a coherent contribution, described by the most-crossed
diagrams, over an incoherent contribution represented by the ladder diagrams.
The parameter \( \lambda  \) plays the role of an ``absorption'' term for
the coherent contribution only, and is therefore responsible for the decrease
of the coherent backscattering cone in a magnetic field. This parameter \( \lambda  \)
can be expressed as the square of a Faraday dephasing angle, which is the product
of a Verdet constant, the magnetic field strength, and a characteristic length
scale for the propagation of light in the medium. Since the scattering mean
free path \( \ell  \) is a natural and experimentally relevant length scale
for coherent backscattering, it is possible to define an effective Verdet constant
\( V_{eff} \) from a relation derived in \cite{john},

\begin{equation}
\label{veff}
\lambda =\frac{4}{3}V^{2}_{eff}B^{2}\ell ^{2}.
\end{equation}
 Equation (\ref{veff}) was obtained by MacKintosh and John \cite{john}, who
considered a situation where the scatterers and the outside medium both have
a Verdet constant \( V_{eff} \) . 

To first order in the magnetic field, the tensor \( d_{ik} \) defined in Eq.
(\ref{C}) must take the form

\begin{equation}
\label{d}
\mathbf{d}(\mathbf{p},\mathbf{q}=0,\mathbf{B})=\mathbf{I}+d_{2}\Phi .
\end{equation}
 where \( d_{2} \) describes persisting polarization effects in diffuse scattering
due to the magnetic field \cite{euro}. For low density \( n \) of particles,
\( d_{2} \) can be determined from the Bethe-Salpether equation

\begin{equation}
\label{BS}
L_{\mathbf{pp}'}^{-}(\mathbf{B})=n\mathbf{T}_{\mathbf{pp}'}(\mathbf{B})
\mathbf{T}^{*}_{\mathbf{pp}'}(-\mathbf{B})+n\sum _{\mathbf{p}''}
\mathbf{T}_{\mathbf{pp}''}(\mathbf{B})\cdot \mathbf{G}(\mathbf{p}'',
\mathbf{B})\cdot \mathbf{G}^{*}(\mathbf{p}'',-\mathbf{B})\cdot \mathbf{T}^{*}_
{\mathbf{pp}''}(-\mathbf{B})\cdot L_{\mathbf{p}''\mathbf{p}'}^{-}(\mathbf{B}),
\end{equation}
 where \( \mathbf{G}(\mathbf{p},\mathbf{B}) \) denotes the ensemble averaged
Dyson Green's tensor (see Appendix A for details of notations), and the asterisk
denotes hermitian conjugation in polarization space. This equation is slightly
more complicated than the one for \( L_{\mathbf{pp}'}^{+} \) \cite{euro} because
it involves both \( \mathbf{B} \) and \( -\mathbf{B} \) due to the definition
(\ref{C}). Inserting Eqs. (\ref{C}) and (\ref{d}) into Eq. (\ref{BS}) and
expanding to first and second order in the magnetic field using Eq. (\ref{Tkkp}),
fixes \( d_{2} \) and \( \lambda  \) rigorously. The final result reads 

\begin{equation}
\label{d2}
d_{2}=i\left( \frac{-A_{2}}{1-\left\langle \cos \theta \right\rangle _{p}}-\eta \right) ,
\end{equation}

\begin{equation}
\label{lambda}
\lambda =\frac{1}{3}\left( A^{2}_{1}+\frac{A^{2}_{2}}{1-\left\langle 
\cos \theta \right\rangle _{p}}\right) ,
\end{equation}
 with 
\begin{equation}
\label{A1}
A^{2}_{1}=\frac{1}{x^{2}Q_{scatt}}\int ^{1}_{-1}d\mu \left[ 2\frac{|R_{1}
|^{2}+|R_{2}|^{2}-2\Re e(R_{1}\overline{R_{2}})\mu }{1-\mu ^{2}}+|Q_{1}|^{2}+
|Q_{2}|^{2}\right] ,
\end{equation}

\begin{equation}
\label{A2}
A_{2}=\frac{1}{x^{2}Q_{scatt}}\int ^{1}_{-1}\Re e(R_{1}S_{2}+R_{2}S_{1})d\mu -\eta .
\end{equation}

Equations (\ref{d2}-\ref{A2}) contain the main results of this paper and will
be discussed in sections \ref{Sec: dephasing} to \ref{Sec:shift}. The factor
\( 1/(1-\left\langle \cos \theta \right\rangle _{p}) \) in Eqs. (\ref{d2})
and (\ref{lambda}) is related to a depolarization length which will be introduced
in the next section. In equation (\ref{lambda}), \( \lambda  \) contains two
terms: the first one is seen to dominate for finite size particles (\( x\gg 1) \)
whereas the second one prevails for small particles (\( x\ll 1 \)). Sections
\ref{Sec: dephasing} and \ref{Sec:dephasing:Mie} will discuss these two contributions
respectively. Far from resonances, the first term corresponds to the Faraday
rotation of the wave inside the scatterer, whereas the second represents the
Faraday rotation between two scattering events. Although the medium outside
the scatterers is not Faraday active, the Faraday rotation from one scatterer
to the next can be defined in the framework of an effective medium theory using
the parameter \( \eta  \), which will be defined in section \ref{Sec: dephasing}.

\section{Depolarization length}

\begin{figure}
{\par\centering \resizebox*{9cm}{9cm}{\rotatebox{-90}{\includegraphics{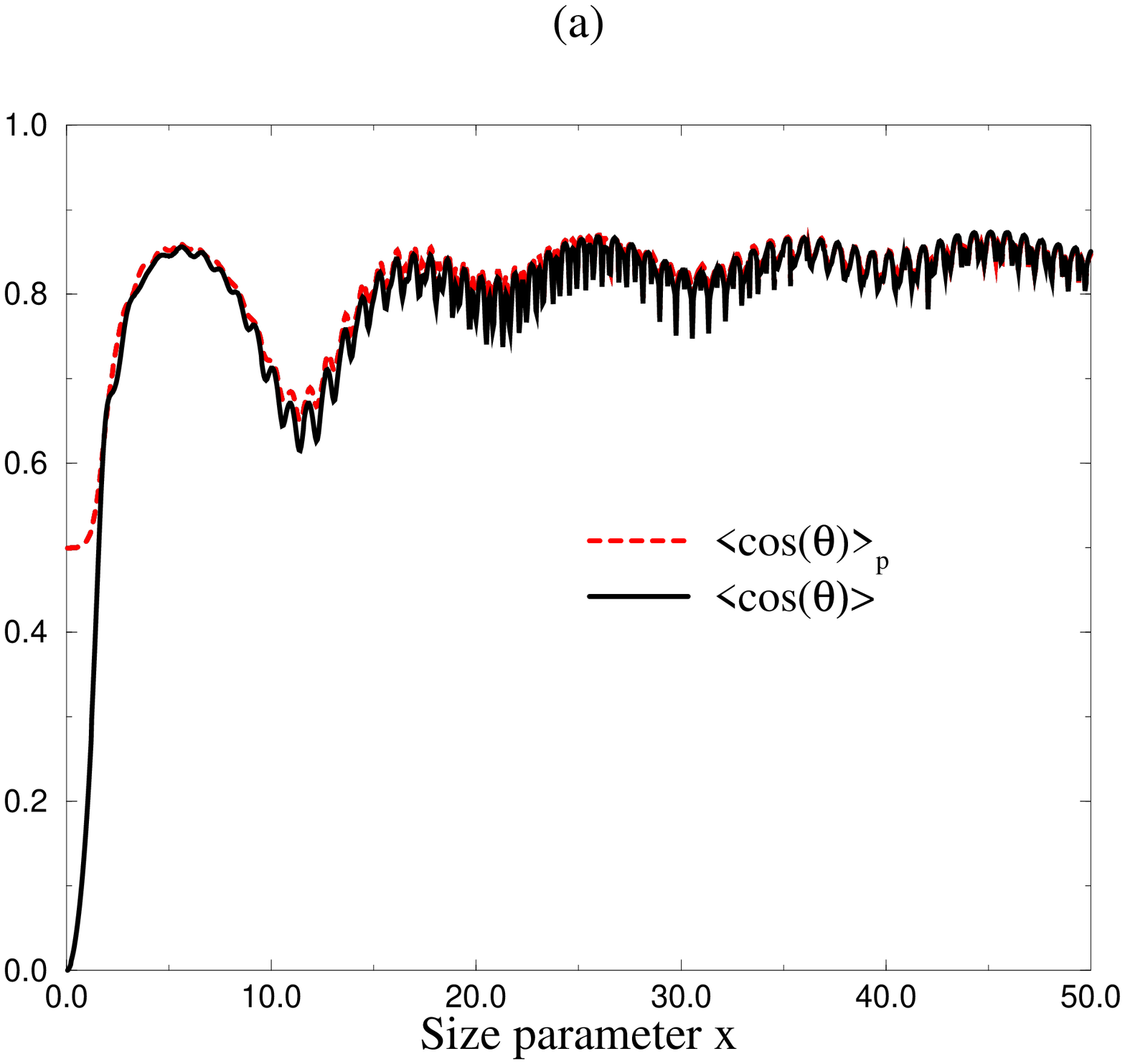}}} 
\resizebox*{9cm}{9cm}{\rotatebox{-90}{\includegraphics{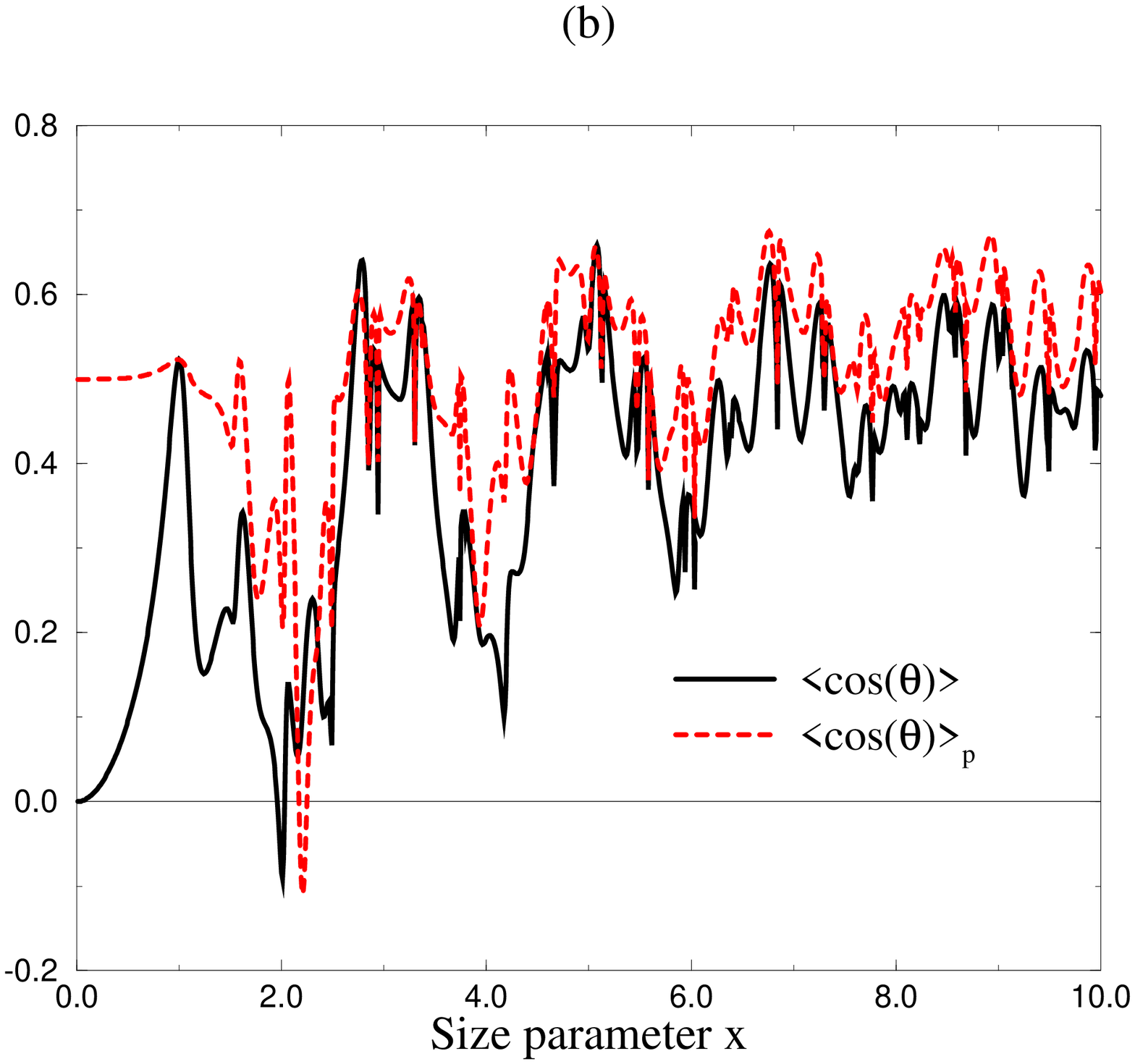}}} \par}

\caption{Asymmetry parameters \protect\( \left\langle \cos \theta \right\rangle \protect \)
and \protect\( \left\langle \cos \theta \right\rangle _{p}\protect \) as a
function of the size parameter \protect\( x\protect \) for a relative index
of refraction \protect\( m=1.33\protect \) (a) and \protect\( m=2.73\protect \)
(b)\label{Depol}. }
\end{figure}

In this section, the factor \( 1/(1-\left\langle \cos \theta \right\rangle _{p}) \),
which tends to amplify the second term of Eq. (\ref{lambda}) is defined. The
term \( A^{2}_{2} \) is easily seen to dominate for small particles. The total
cross-section of one particle is given by \cite{hulst},
\begin{equation}
\label{qscatt}
Q_{scatt}=\frac{1}{x^{2}}\int ^{1}_{-1}([S_{1}|^{2}+|S_{2}|^{2})d\mu ,
\end{equation}
with \( \mu =\cos \theta  \). The transport mean free path is defined as \( \ell 
^{*}=\ell /(1-\left\langle \cos \theta \right\rangle ), \)
where the asymmetry parameter \( \left\langle \cos \theta \right\rangle  \)
is given by
\[
\left\langle \cos \theta \right\rangle =\frac{1}{x^{2}Q_{scatt}}\int ^{1}_{-1}
([S_{1}|^{2}+|S_{2}|^{2})\mu d\mu .\]
 Likewise, a depolarization length can be defined as \( \ell _{dep}=\ell 
 /(1-\left\langle \cos \theta \right\rangle _{p}), \)
where 
\[
\left\langle \cos \theta \right\rangle _{p}=\frac{1}{x^{2}Q_{scatt}}
\int ^{1}_{-1}2\Re e(S_{1}\overline{S_{2}})\mu d\mu .\]
 Rayleigh scattering has a forward-backward symmetry so that \( \left\langle 
 \cos \theta \right\rangle =0 \).
However, scattering is not isotropic due to the polarization and one can easily
show that for Rayleigh scatterers \( \left\langle \cos \theta \right\rangle _{p}=0.5 \).
In the limit of large forward scattering and for \( m=1.33 \), both asymmetry
parameters, \( \left\langle \cos \theta \right\rangle  \) and \( \left\langle 
\cos \theta \right\rangle _{p} \)
tend towards a limit close to \( 0.85 \), as shown in Fig. (\ref{Depol}a).
In the forward direction the differences between the two states of polarization
vanish since \( S_{1}(0)=S_{2}(0). \) 

The well-known oscillations and ripple structure \cite{hulst} of the asymmetry
parameter \( \left\langle \cos \theta \right\rangle  \) are also present in
\( \left\langle \cos \theta \right\rangle _{p} \). As shown in Fig. (\ref{Depol}b),
for a relative high value of the relative index of refraction (\( m=2.73 \)
corresponding to TiO\( _{2} \) ), the asymmetry parameters \( \left\langle \cos 
\theta \right\rangle  \)
and \( \left\langle \cos \theta \right\rangle _{p} \) may take negative values,
which can be seen in this particular case near \( x=2. \) In this very particular
case, where the scattering is essentially in the backward direction, the characteristic
length for the loss of the polarization \( \ell _{dep} \) can be smaller than
the characteristic length for the loss of the phase in multiple light scattering,
which is \( \ell  \).

\section{Faraday rotation for multiple Rayleigh scattering }

\label{Sec: dephasing}Equation (\ref{d2}) for \( d_{2} \) and the second
term in Eq. (\ref{lambda}) can be understood from an ``effective medium''
theory, valid for Rayleigh scatterers. The real part of the forward scattering
amplitude (\ref{tkk}) is associated with the Faraday effect and the imaginary
part with magneto-dichroism (\emph{i.e} different absorption for different circular
polarization) of an ensemble of Faraday-active scatterers. For a dilute system,
the antisymmetric part of this effective refractive index \( \varepsilon _{a} \)
is defined as

\begin{equation}
\label{epsa}
\varepsilon _{a}=-\frac{2\pi n}{\omega ^{3}}R_{1}(0).
\end{equation}
 For the real part of this effective refractive index, the dimensionless parameter
\( \eta  \) is defined by
\begin{equation}
\label{eta}
\eta =\Re e(\varepsilon _{a})k\ell .
\end{equation}
 Equations (\ref{d2}) and (\ref{A2}) involve this parameter \( \eta  \),
which represents a characteristic phase in multiple scattering, due to the Faraday
effect in the effective medium accumulated over a distance \( \ell  \). An
ensemble of Rayleigh scatterers (for which the electromagnetic field changes
only slightly on a scale comparable to the particle size) has the finite value
\begin{equation}
\label{epsar}
\varepsilon _{a}=\frac{9f\varepsilon _{F}}{(m^{2}+2)^{2}}.
\end{equation}
 The Faraday effect of a composite material made of particles smaller than the
wavelength, and of different shape (spherical, needle like, plate like) was
discussed in Ref. \cite{xia}, using a more general version of the effective
medium approximation (not limited to dilute samples).

For Rayleigh scatterers (\( x\rightarrow 0 \)), one obtains

\[
d_{2}=\frac{9i\varepsilon _{F}}{4(m^{2}-1)^{2}x^{3}},\]
\[
\lambda =\frac{27\varepsilon _{F}^{2}}{8(m^{2}-1)^{4}x^{6}},\]
 so that, by definition (\ref{veff}),

\begin{equation}
\label{rayleigh}
\frac{V_{eff}}{fV_{0}}=\frac{9\sqrt{2}m}{(m^{2}+2)^{2}}.
\end{equation}

Apart from a factor depending on the index of refraction, the effective Verdet
constant is found to be the product of the volume fraction of the particles
by their Verdet constant. As noticed before \cite{bart}, one finds a factor
of \( \sqrt{2} \) more than expected on the basis of the effective medium approach
of Eq. (\ref{epsar}) if \( V_{eff} \) would have defined by \( \varepsilon _{a}=
2V_{eff}B/\omega  \).
This discrepancy is due to the denominator in Eq. (\ref{lambda}), \( 1-\left\langle
 \cos \theta \right\rangle _{p}=1/2 \)
for Rayleigh particles.

\begin{figure}
{\par\centering \resizebox*{9cm}{9cm}{\rotatebox{-90}{\includegraphics{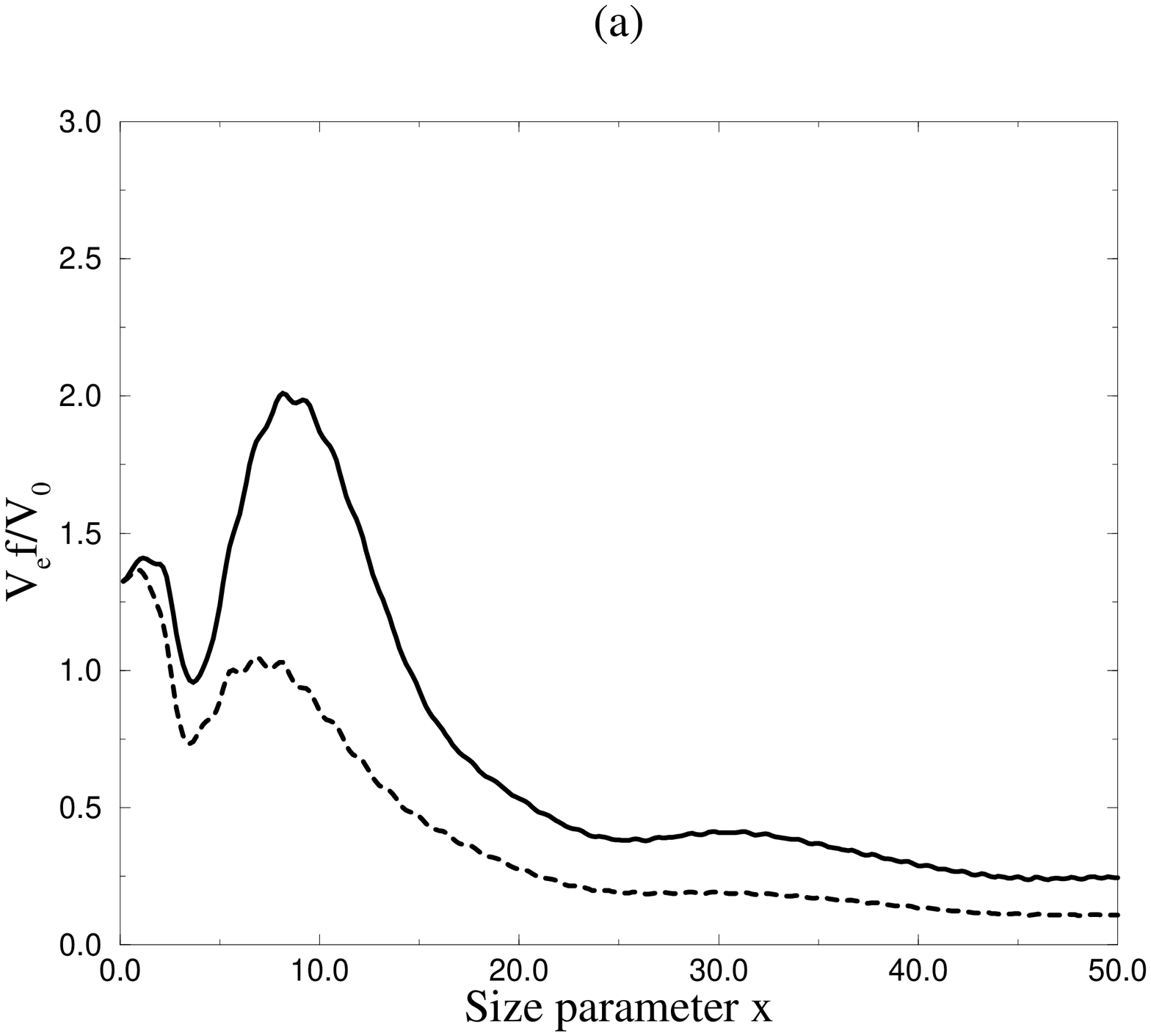}}} 
\resizebox*{9cm}{9cm}{\rotatebox{-90}{\includegraphics{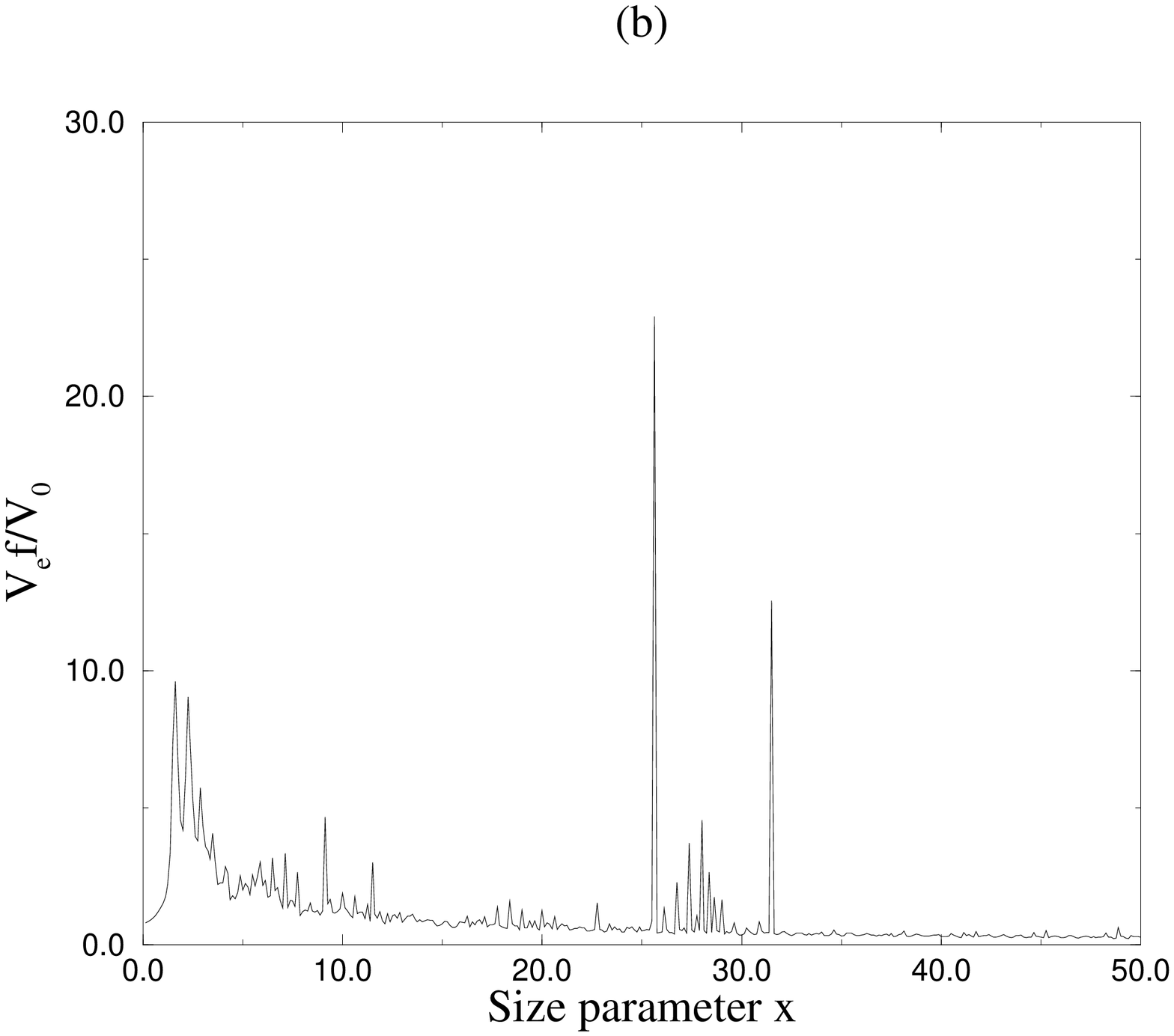}}} \par}

\caption{Plot of the effective Verdet constant \protect\( \frac{V_{eff}}{fV_{0}}\protect \)
as a function of the size parameter \protect\( x\protect \). (a) The solid
line represents the solution of Eqs. (\ref{d2}-\ref{A2}), containing contributions
from all the rays incident on the sphere, whereas the dashed line corresponds
to the contribution of only the first partial wave \protect\( J=1\protect \)
(the ray with central impact in geometrical optics). The scattering medium has
an index of refraction of \protect\( 1.7\protect \), the value in the experiment
of Erbacher et \emph{al..} (b) The same plot for an index of refraction of \protect\( 2.73
\protect \)
for which resonances are clearly visible. The host medium is in both cases glycerol
(of index of refraction \protect\( 1.47\protect \)). A general decrease of
the effective Verdet constant for increasing size parameter \protect\( x\protect \)
can be observed in both plots.\label{fig:verdet}}
\end{figure}

From the experimental parameters described in the experiments of Erbacher \emph{et
al.} (a relative index of refraction of \( m=1.15 \)), the estimate using Eq.
(\ref{rayleigh}) is
\begin{equation}
\label{lim_verdet}
\frac{V_{eff}}{fV_{0}}\approx 1.32.
\end{equation}
 This value is to be compared with the experimental value of \( 1.55\pm 0.15 \)
\cite{lenke}. The proximity of the two values probably explains the success
of previous theories based on Rayleigh scatterers, although this experiment
dealt with Mie particles. In this experiment, the maximum of the distribution
of the size parameters was roughly estimated at \( x\simeq 23 \) but the width
of the distribution was very broad. Using the parameters of Erbacher \emph{et
al.,} our Mie theory reproduces the limit of Eq. (\ref{lim_verdet}) for \( x=0 \),
but predicts a value of only \( 0.4 \) at \( x\simeq 23 \). The solid line
in figure \ref{fig:verdet}a represents the effective Verdet constant as a function
of the size parameter \( x \), for the same relative index of refraction used
in the experiment of Erbacher et \emph{al.}. One can clearly see in this figure
that a distribution of large spheres of size parameter \( x \) of the order
of \( 20 \) or higher can not explain the experimental result. However if the
size distribution of the scatterers was rather centered around a size parameter
of roughly \( 10 \), the experimental value of the effective Verdet constant
could possibly be recovered from this Mie theory. In figure \ref{fig:verdet}b,
an higher value of the relative index was chosen. In this case, the effective
Verdet constant is seen to be enhanced by resonances, which will be the subject
of the section \ref{Sec:resonance}.

\section{Faraday rotation for multiple Mie scattering }

\label{Sec:dephasing:Mie}The first term in Eq. (\ref{lambda}) originates from
the Faraday effect inside the scatterers and is the main contribution in the
Mie regime of \( x\gg 1 \). Resonances will be discussed in section \ref{Sec:resonance}.
Using the definition of Eq. (\ref{veff}), the effective Verdet constant of
Mie particles is written as

\begin{equation}
\label{Mie}
\frac{V_{eff}}{fV_{0}}=\frac{3m\, Q_{scatt}\sqrt{3\lambda }}{4x\varepsilon _{F}}.
\end{equation}

When the size parameter obeys \( x\gg 1 \) and \( y=mx\gg 1 \), the scattering
can be interpreted in terms of geometrical optics. In geometrical optics, rays
incident on the sphere are considered rather than plane waves. The ray with
central impact is characterized in Mie theory by \( J=1 \). In figure \ref{fig:verdet}a,
the effective Verdet constant, that contains contributions from all the rays,
is plotted as a solid line, with respect to the size parameter \( x \). The
dashed line in this figure represents the separated contribution of the ray
with central impact \( J=1 \) in the effective Verdet constant. The two curves
merge for \( x=0 \), the Rayleigh limit, but deviate from each other for larger
size parameter. For a given value of the size parameter, it can be noted that
the ray with central impact already represents a significant contribution to
the effective Verdet constant. In addition to this, the effective Verdet constant
is seen to decrease for increasing size parameter \( x \). This is an important
observation, because this means that Mie scatterers of large size (typically
\( x>20 \)) are less efficient than Rayleigh scatterers in suppressing the
coherent backscattering cone.

\section{Mie resonances}

\label{Sec:resonance} Let us first recall some results for resonant Rayleigh
scattering, for which the resonance behavior is analogous to the resonance of
a two-level atom in atomic physics. For resonant Rayleigh scatterers the effective
Verdet constant is related to the ``path length'' \( L_{path} \) of the wave
inside the particle. In fact, the path length, the dwell time of the light and
the total electromagnetic energy stored inside the scatterer have been seen
to be proportional. The path length can be defined as (see formula (2-22) P17
of Ref. \cite{ishimaru}),

\begin{equation}
\label{lp_def}
L_{path}=\lim _{m_{i}\rightarrow 0}\frac{Q_{abs}}{\omega m_{i}},
\end{equation}
where \( Q_{abs} \) is the absorption cross-section and \( m_{i} \) the imaginary
part of the index of refraction. The physical idea behind this definition is
that the longer the path of the light is in the particle, the more the light
will suffer from absorption. For resonant Rayleigh particles, the relation to
the effective Verdet constant is obtained from Eqs. (\ref{rayleigh},\ref{lp_def}),

\begin{equation}
\label{lpath}
\frac{V_{eff}}{fV_{0}}=\frac{4m_{r}\sqrt{2}}{3}\frac{L_{path}}{a}.
\end{equation}
 At resonance, the path length can exceed the size of the scatterer, which means
that the effective Verdet constant should be strongly enhanced by resonant scattering.
Alternatively, one may relate the path length to the time spent by the wave
in the medium, which means that the Faraday rotation is in some sense a \emph{Larmor
clock}, measuring this time \cite{landauer}. 

\begin{figure}
{\par\centering \resizebox*{9cm}{9cm}{\rotatebox{-90}{\includegraphics{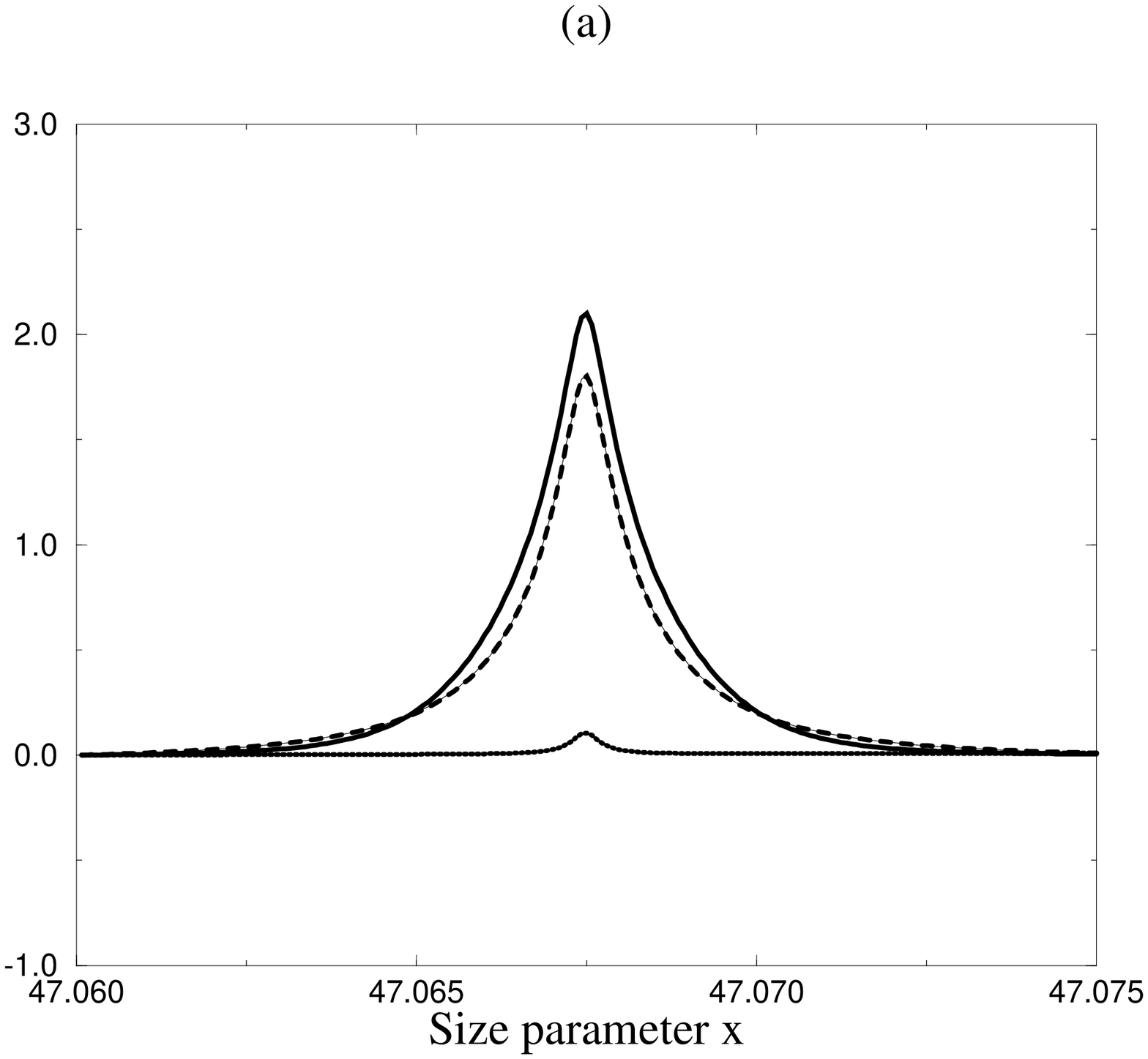}}} 
\resizebox*{9cm}{9cm}{\rotatebox{-90}{\includegraphics{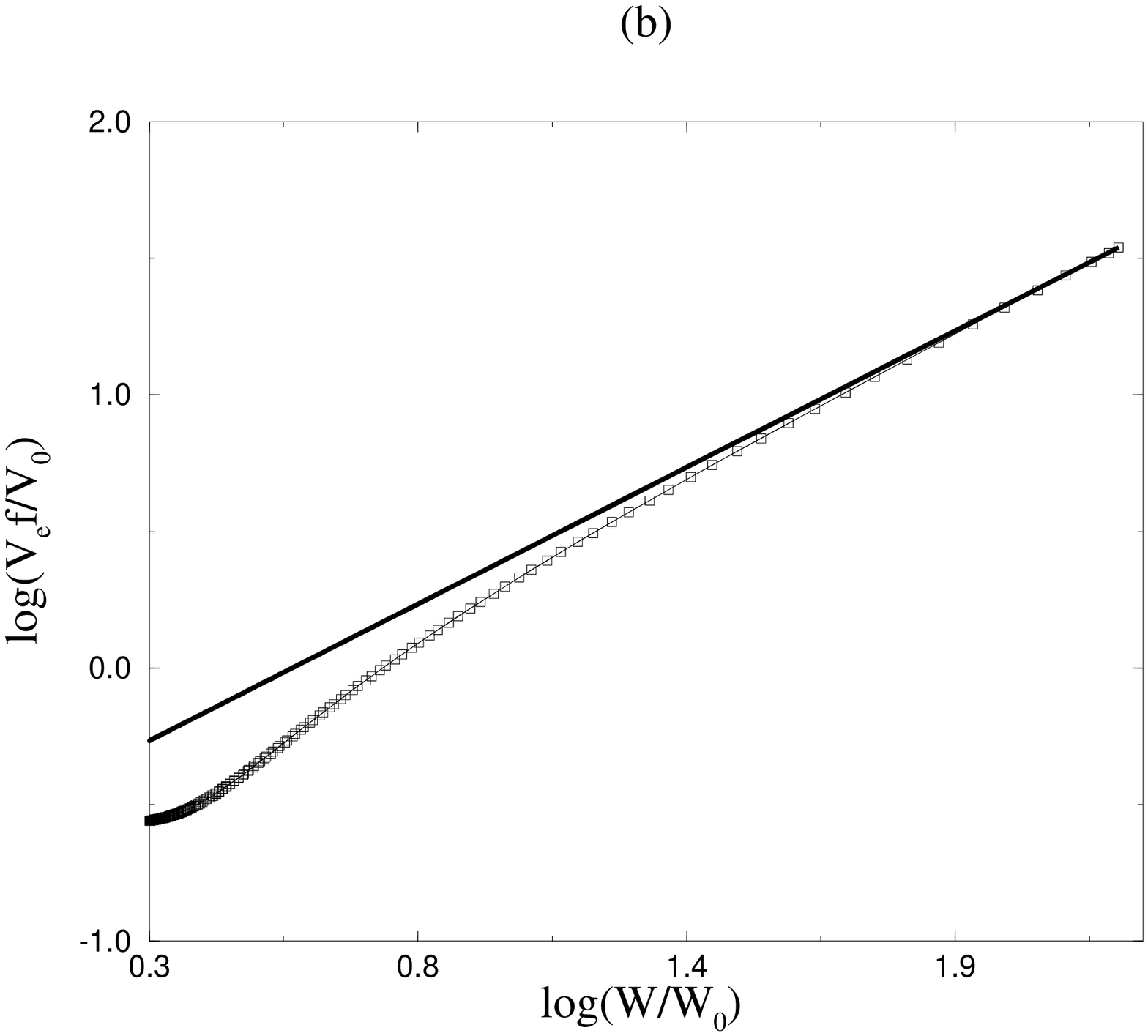}}} \par}

\caption{Near a particular resonance curve of a water sphere of index 
\protect\( m=1.334-1.5\cdot 10^{-9}i\protect \)
the following curves have been plotted: (a) Respectively from the upper part
of the figure to the bottom: a plot of \protect\( \log (V_{eff}/(fV_{0}))\protect \)
(solid), \protect\( \log W\protect \) (dashed), and the scattering cross-section
\protect\( Q_{scatt}\protect \) (dotted) as a function of the size parameter
\protect\( x\protect \). (b) Close to this particular resonance, 
\protect\( \log (V_{eff}/(fV_{0}))\protect \)
is shown against the total electromagnetic energy \protect\( W\protect \) (dots),
to be compared with a line of slope one (solid), the prediction of Eq. (\ref{prop}).
\label{resonance}}
\end{figure}

The question is whether resonant enhancement of Faraday rotation occurs in resonant
Mie scattering. For Mie resonances the increase in path length is well related
to the change of the electromagnetic energy within the scatterers with respect
to the surrounding. The total time-averaged electromagnetic energy inside the
sphere is denoted by \( W \) and \( W_{0} \) represents this energy for the
incident plane wave. For weak absorption, \( m_{i}\ll m_{r}, \) the electromagnetic
energy \( W \) can be approximated by \cite{bott} 
\begin{equation}
\label{lpath2}
\frac{W}{W_{0}}\simeq \frac{3m_{r}Q_{abs}}{8xm_{i}}=\frac{3m_{r}L_{path}}{a}.
\end{equation}
 where Eq. (\ref{lp_def}) was applied to obtain the last equality. This relation
is exact for scalar waves and is a very good approximation for vector waves.
It is even an excellent approximation in the vicinity of resonances were the
deviations between the exact solution and its approximation are the largest.
In the domain of \( J\simeq x \), several resonances take place in the Mie
coefficients \( a_{J} \) and \( b_{J} \). These resonances are well separated
and can be numbered by an additional integer \( k \) the order of the resonance
\cite{chylek}. Near one electric Mie resonance of a specified order, the denominators
of the Mie coefficients \( a_{J} \) and \( c_{J} \), which are identical,
are close to zero. From Eqs. (\ref{lambda},\ref{A1}), one finds \( \sqrt{\lambda }
\sim |c_{J}|^{2}/|a_{J}|. \)
By Eqs. (\ref{Mie}) and (\ref{lpath2}), this implies

\begin{equation}
\label{prop}
\frac{V_{eff}}{fV_{0}}\sim |c_{J}|^{2}|a_{J}|\sim \frac{W}{W_{0}}|a_{J}|\sim \frac{W}{W_{0}}.
\end{equation}
 In the last equality the role of \( |a_{J}| \) is dominated by \( |c_{J}|^{2} \)
at resonance, since resonances in the scattering cross-section \( Q_{scatt} \)
are much less significant than resonances in \( W \). Indeed, in Fig. (\ref{resonance}a),
the lower curve represents \( Q_{scatt} \), on normal scale, near one resonance
of a water sphere. It is much below the curves of \( W \) or \( V_{eff}/(fV_{0}) \)
which are even plotted on a logarithmic scale. From Eq. (\ref{prop}), the effective
Verdet constant is expected to be simply proportional to the electromagnetic
energy \( W \) (or equivalently to \( L_{path} \)). This generalizes the result
for Rayleigh scatterers in Eq. (\ref{lpath}). The numerical verification of
Eq. (\ref{prop}) can be deduced from the double logarithmic plot of \( V_{eff}/(fV_{0}) \)
against the total electromagnetic energy \( W \) in Fig. (\ref{resonance}b). 

The proportionality of the Verdet constant and the total electromagnetic energy
(or equivalently \( L_{path} \)) has been derived for the particular case where
the path of the light is confined along the same line (1D problem as in a Fabry-Perot
configuration for instance \cite{josa}). In this case, the cumulative character
of the Faraday rotation with respect to the path length leads to an experimentally
observed enhancement of the Faraday rotation \cite{fp}. Eq. (\ref{prop}) applies
to any resonant impact and shows that the possible occurrence of spin flips
in Mie scattering, as suggested in Ref. \cite{lenke}, does not affect the behavior
of the effective Verdet constant near resonances and that the Faraday rotation
still accumulates along the path as in the 1D case. This is consistent with
the observation made in section \ref{Sec:dephasing:Mie}, that the ray with
central impact had an important role for interpreting the Faraday rotation for
multiple Mie scattering. 

In conclusion, like for resonant Rayleigh particles, a strong correlation between
the effective Verdet constant and the stored energy inside the sphere was found
for resonant Mie particles. The general behavior and proportionality in the
vicinity of a resonance is apparently universal.

\section{Shift of the intensity profile of the coherent backscattering cone in a magnetic
field }

\label{Sec:shift}A magnetic field can be expected to introduce some anisotropy
in the light intensity profile of the cone. In this section, the form taken
by this anisotropy is investigated by taking care of the selection rules imposed
for the polarization in reflection of a semi-infinite system of Mie scatterers.
This analysis is restricted to linear corrections in the magnetic field, so
that the parameter \( \lambda  \) discussed in the former section, quadratic
in the field, will no longer appear. 

This approach is based on an improved version of the scalar diffusion approximation.
The ladder propagator at point \( \mathbf{r}=\{\mathbf{r}_{\bot },z\} \) for
a source at \( \mathbf{r}'=\{\mathbf{r}_{\bot }',z'\} \) in a semi-infinite
medium is denoted by \( \rho (\mathbf{r},\mathbf{r}') \). The \( z \)-axis
is directed along the normal of the sample, and \( \mathbf{r}_{\bot } \) and
\( \mathbf{r}_{\bot }' \) are vectors perpendicular to the \( z \)-axis. Because
of translational invariance in the plane of the sample, the ladder propagator
only depends on \( \mathbf{r}_{\bot }-\mathbf{r}_{\bot }' \). The two-dimensional
Fourier transform of \( \rho (\mathbf{r},\mathbf{r}') \) with respect to \( \mathbf{r}_
{\bot }-\mathbf{r}_{\bot }' \)
is denoted by \( \widetilde{\rho }(\mathbf{q},z,z') \). The ladder propagator
\( \rho (\mathbf{r},\mathbf{r}') \) obeys the following diffusion equation
\begin{equation}
\label{diffusion}
(-\nabla ^{2}+\frac{1}{L_{a}^{2}})\rho (\mathbf{r},\mathbf{r}')=\delta (\mathbf{r}-
\mathbf{r}'),
\end{equation}
 with the radiative boundary condition \cite{vanderMark}

\begin{equation}
\label{BC}
\forall \mathbf{r}_{\bot },\forall \mathbf{r}_{\bot }',\forall z>0,\, \, \, \, \, \, 
\, \, \, \, \, \, \, \, \, \, \, \, \, \, \rho (\{\mathbf{r}_{\bot },z\},\{\mathbf{r}_
{\bot }',z'=-z_{0}\})=0.
\end{equation}

The trapping plane is located at a distance \( z_{0}=2\ell ^{*}/3 \) outside
the sample, and \( L_{a} \) is the absorption length for the light intensity.
We assume that the first and the last scattering events take place one transport
mean free path \( \ell ^{*} \) away from the boundary, in the directions specified
by the incoming wavevector \( \mathbf{p} \) and the outgoing wavevector \( \mathbf{p}' \).
This allows to calculate the contribution of the ladder diagrams \( L^{+}_{ijkl}(\mathbf{p},
\mathbf{p}',\mathbf{q},\mathbf{B}) \),
the so-called incoherent contribution to the coherent backscattering, directly
from \( \widetilde{\rho }(\mathbf{q}=0,z-\ell ^{*}\hat{p}_{z},z'+\ell ^{*}\hat{p}_{z}') \).
Similarly, the coherent contribution is obtained from the most-crossed diagrams
\( C^{+}_{ijlk}(\mathbf{p},\mathbf{p}',\mathbf{q},\mathbf{B}) \), which are
derived from \( L^{-}_{ijkl}(\mathbf{p},\mathbf{p}',\mathbf{q},\mathbf{B}) \)
using Eq. (\ref{CL}). These derivations are detailed out in Appendix A. The
coherent contribution in an helicity basis and at reflection (\( z=0,z'=0 \))
reads,

\begin{equation}
\label{fin}
C^{+}_{\sigma \sigma '}=\widetilde{\rho }(\mathbf{p}_{\bot }+\mathbf{p}_{\bot }',-\ell 
^{*}\hat{p}_{z},\ell ^{*}\hat{p}_{z}')\times \delta _{\sigma \sigma '}\times (1+2b_{1}
\omega \ell \det (\hat{\mathbf{p}}_{\bot }',\hat{\mathbf{p}}_{\bot },\hat{\mathbf{B}})-2
\sigma b_{2}\omega \ell \hat{\mathbf{B}}\cdot (\hat{\mathbf{p}}_{\bot }+\hat{
\mathbf{p}}_{\bot }')).
\end{equation}
where the Kronecker symbol \( \delta _{\sigma \sigma '} \) guarantees conservation
of helicity \( \sigma (\hat{\mathbf{p}})=\sigma '(\hat{\mathbf{p}}') \), and
\( b_{1} \) and \( b_{2} \) are real-valued coefficients to be determined.
Most experiments on coherent backscattering were done in the helicity-conserving
channel, which has the advantage of having a maximal enhancement factor (since
the contribution from single scattering vanishes in this case) and of having
an isotropic line shape. Eq. (\ref{fin}) states that the magnetic field modifies
the cone exactly in this channel, in agreement with previous work \cite{john}.
Even when a magnetic field is present, no coherent backscattering is found in
the opposite-helicity channel, at least according to the present diffusion approximation. 

Only the components of the magnetic field along the slab contribute in the r.h.s
of Eq. (\ref{fin}). When the field is perpendicular to the slab, the decrease
of the enhancement factor, described by \( \lambda  \), is the sole impact
of the magnetic field and is not included in the present approximation. When
the magnetic field is in the plane of the slab, two corrections show up. The
first one, proportional to \( b_{1} \), is magneto-transverse, since it produces
a shift of the intensity profile of the cone in the plane of the slab, normal
to the magnetic field. This correction is independent of the state of helicity
of the light. The second correction in Eq. (\ref{fin}) proportional to \( b_{2} \)
does depend on the helicity \( \sigma  \). It produces a shift of the intensity
profile of the cone in the direction of the magnetic field in the plane of the
slab, quite similar to the correction induced by the magnetic field in the group
velocity \cite{landau2}. 

The coefficient \( b_{1} \) can be calculated independently from \( b_{2} \),
in a way exactly analogous to the calculation of \( a_{1} \) in Ref. \cite{euro}
responsible for the \emph{Photonic Hall Effect} (PHE). The result is

\begin{equation}
\label{b1}
b_{1}=\frac{1}{(1-\left\langle \cos \theta \right\rangle )^{2}}\frac{\int ^{1}_{-1}d\cos
 \theta \sin \theta \, \sum _{\sigma \sigma '}\Im m\left( T^{0}_{\sigma \sigma '}(\theta )
 \overline{T^{1}_{\sigma \sigma '}}(\hat{\mathbf{B}}=\hat{\mathbf{g}},\theta )\right) }
 {2\int ^{1}_{-1}d\cos \theta \sum _{\sigma \sigma '}|T^{0}_{\sigma \sigma '}(\theta )|^{2}},
\end{equation}
where the magneto T-matrix \( T^{1}_{\sigma \sigma '}(\hat{\mathbf{B}}=\hat{\mathbf{g}})
(\theta ) \)
was introduced in Eq. (\ref{tbg}). Note that the imaginary part in Eq. (\ref{b1})
discriminates \( b_{1} \) from the parameter \( a_{1} \) responsible for the
PHE, in which the \emph{real} part figures. For Rayleigh scattering, one can
readily prove that the parameters \( b_{1} \) and \( b_{2} \) both vanish.
The calculation of \( b_{2} \) for Mie scattering is very complicated and is
beyond the scope of this paper.

The expected modification of the lines of equal enhancement factor due to the
magnetic field is now investigated. Equation (\ref{fin}) translates into a
CBS line shape

\begin{equation}
\label{fin2}
E(\mu ,\varphi )=1+C(\mu )(1+2b_{1}\mu \sin \varphi -2b_{2}\sigma \mu \cos \varphi ).
\end{equation}
 The enhancement factor, in the helicity-conserving channel, is denoted by \( E \)
for a state of helicity \( \sigma  \). The dimensionless parameter \( \mu =\omega \ell \theta  \)
and \( C(\mu ) \) the well-documented line shape of the cone without applied
magnetic field were introduced \cite{bart}. The azimuthal angle between the
projection of the outgoing wave vector into the plane of the slab and the magnetic
field direction, which has been chosen along the \( x \)-axis, was denoted
by \( \varphi  \). For simplicity, only the magneto-transverse correction proportional
to \( b_{1} \) will be considered here, so that \( b_{2}=0 \). This case corresponds
to unpolarized incident light, for which the term proportional to \( b_{2} \)
in Eq. (\ref{fin2}) vanishes. The pattern of the lines of equal enhancement
factor associated with the \( b_{2} \)-correction alone is the same as the
one of the transverse correction after a rotation of angle \( \pi /2 \) about
the \( x \)-axis. For a typical experiment \cite{nature} with CeF\( _{3} \)
particles of approximate radius of \( 2\mu  \)m at room temperature, with a
wavelength \( \lambda =457 \)nm, Eq. (\ref{b1}) leads to the estimation \( b_{1}\simeq 1.8\cdot 
10^{-2}. \)
The experimentally measured mean free path for a volume fraction of \( f=0.1 \)
is \( \ell ^{*}\simeq 90\mu  \)m. Eq. (\ref{fin2}) is valid for \( b_{1}\mu \ll 1 \),
which means that our approach is limited to the angular domain \( |\theta |\leq 0.8 \)rad.
The equation for the lines of constant enhancement factor \( E_{0} \) in the
absence of magnetic field is \( E_{0}=1+C(\mu ^{0}) \), independent of the
azimuthal angle \( \varphi  \). The first order correction in the magnetic
field is separated by writing \( \mu =\mu ^{0}+\mu ^{1} \). Equation (\ref{fin2})
gives 
\begin{equation}
\label{mu}
\mu ^{1}=-2b_{1}\frac{C(\mu ^{0})}{C'(\mu ^{0})}\mu ^{0}\sin \varphi .
\end{equation}

\begin{figure}
{\par\centering \resizebox*{9cm}{9cm}{\includegraphics{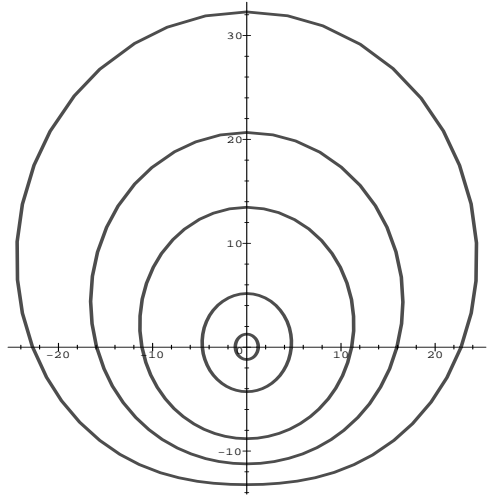}} \par}

\caption{Polar plot of the lines of constant enhancement factor of the coherent backscattering
cone in a magnetic field, in the helicity-conserving channel, for \protect\(
 b_{1}=1.8\cdot 10^{-2}\protect \).
The magnetic field is along the horizontal axis on this graph. The graduations
of the axis represent the dimensionless parameter \protect\( \mu =\omega \ell \theta 
\protect \).
For instance, a graduation of \protect\( 10\protect \) corresponds to an angle
of \protect\( 145\protect \)mrad (for the mean free path \protect\( \ell ^{*}\simeq 90\mu 
\protect \)m,
and the wavelength \protect\( \lambda =457\protect \)nm mentioned in the text).
Without magnetic field or for a magnetic field perpendicular to the slab, these
lines of constant enhancement factor would have been circles. \label{Fig_lineshape}}
\end{figure}

In Fig. (\ref{Fig_lineshape}), the polar diagram of the lines of constant enhancement
factor is shown for \( b_{1}=1.8\cdot 10^{-2}. \) As apparent from Eqs. (\ref{fin2},\ref{mu}),
the distortion of the lines should increase away from backscattering (at exact
backscattering there is no modification at all to first order in the magnetic
field). As a consequence, with the value of \( b_{1} \) given above only a
modification of the line shape in the wings might be observed. Although the
condition \( b_{1}\mu \ll 1 \) limits the domain of validity of the approximation
there should nevertheless be a sufficiently broad angular range, in the wings
of the cone, where the magnetic corrections could be visible. In figure \ref{Fig_lineshape}
the condition of validity of the approximation has been satisfied. 

Another condition of validity lies in the use of the diffusion approximation.
This approximation predicts a \( 1/\mu ^{2} \) behavior for the line shape
of the cone in the wings, which is actually a wrong result. In the wings, the
contribution of lowest orders of scattering is dominant and not properly taken
into account in the diffusion approximation. Using an exact theory, Gorodnichev
proved that the outcome is a \( 1/\mu  \) dependence \cite{gorodnichev}. His
result has been derived only for point-like scatterers but it should also be
valid for Mie scatterers when the mean free path is much larger than the wavelength.
In any case, the magnetic correction in Eq. (\ref{mu}) depends only on the
logarithmic derivative of \( C(\mu ^{0}) \), which should change only by a
factor of two if the actual law is \( 1/\mu  \) or \( 1/\mu ^{2} \), the general
pattern of the lines of equal enhancement factor being not modified. Therefore,
the shift of the center of mass of the light intensity profile which was calculated,
should be fairly robust with respect to the exact form of \( C(\mu ^{0}) \).

\section{Conclusion}

This paper describes two modifications on the coherent backscattering cone produced
by a magnetic field. The first one, the decrease of the enhancement factor,
depends on the parameter \( \lambda  \) quadratic in the magnetic field, and
was observed experimentally. The second modification is related to the anisotropy
of the light intensity and appears already in linear order of the magnetic field.
Preliminary experiments seem to have reported the possibility of a shift in
the intensity profile \cite{lenke2}. Our analysis applies to spherical scatterers
of any size that are Faraday active. The decrease of the backscattering cone
gets less pronounced as the size of the scatterers increases, whereas the shift
in the intensity profile is only possible with finite size scatterers. As was
surmised in Ref. \cite{lenke}, the effective Verdet constant defined from the
decrease of the cone is enhanced near Mie resonances. The effective Verdet constant
is found to be intimately related to the stored electromagnetic energy \emph{i.e}
the dwell time of the light in the particle. 

We acknowledge R. Lenke for making available his recent experimental work on
the effect of a magnetic field on coherent backscattering light. We thank JJ.
Greffet for stimulating discussions.

\appendix

\section{Derivation of the magnetic corrections to the tensor \( C^{+}_{\sigma \sigma '} \)}

In this appendix, the notations are explained and a demonstration of Eq. (\ref{fin})
is given. The transverse part of the free Green tensor is denoted
\begin{equation}
\label{green}
\mathbf{G}_{0}=\frac{\Delta _{\hat{\mathbf{p}}}}{(\omega /c_{0})^{2}+i\varepsilon -p^{2}},
\end{equation}
 with \( (\Delta _{\hat{\mathbf{p}}})_{ij}=\delta _{ij}-\hat{p}_{i}\hat{p}_{j} \)
the projector upon the space orthogonal to \( \mathbf{p}. \) Similarly, the
hermitian projector on the space orthogonal to \( \mathbf{p} \) for a given
state of helicity \( \sigma  \) is \( P^{\sigma }_{ij}(\hat{\mathbf{p}})=\frac{1}{2}\left( 
(\Delta _{\hat{\mathbf{p}}})_{ij}-i\sigma \varepsilon _{ijk}\hat{p}_{k}\right)  \)
.

Generalizing Eq. (\ref{d}) for finite value of \( \mathbf{q} \) (to first
order in \( \mathbf{q} \) and in the magnetic field \( \mathbf{B} \)) gives
\begin{equation}
\label{dq}
\mathbf{d}(\mathbf{p},\mathbf{q},\mathbf{B})=\mathbf{I}+d_{2}\Phi +\left[ \mathbf{L}
(\mathbf{p},\mathbf{q})-\Gamma ^{-}(\mathbf{p},\mathbf{q},\mathbf{B})\right] \frac{\ell }
{2i\omega },
\end{equation}
 where \( \mathbf{L}(\mathbf{p},\mathbf{q})=2(\mathbf{p}\cdot \mathbf{q})\mathbf{I}-
 \mathbf{pq}-\mathbf{qp} \)
and \( \Gamma ^{-}(\mathbf{p},\mathbf{q},\mathbf{B}) \) are tensors of rank
two, linear in \( \mathbf{q} \), that determine the anisotropy in diffuse scattering.
Without a magnetic field it is well known that \( \Gamma ^{0}(\mathbf{p},\mathbf{q})=
2(\mathbf{p}\cdot \mathbf{q})/(1-\left\langle \cos \theta \right\rangle ). \)
When a magnetic field is present, the first order correction in the magnetic
field is separated as \( \Gamma ^{-}(\mathbf{p},\mathbf{q},\mathbf{B})=\Gamma ^{0}(
\mathbf{p},\mathbf{q})+B\delta \Gamma ^{-}(\mathbf{p},\mathbf{q},\hat{\mathbf{B}}) \).
Because of the symmetry relation of Eq. (\ref{reciprocity}), one has \( \Gamma ^{-}(
\mathbf{p},\mathbf{q},\mathbf{B})=\Gamma ^{-}(\mathbf{p},\mathbf{q},-\mathbf{B})^{*} \).
This implies that \( \delta \Gamma ^{-}(\mathbf{p},\mathbf{q},\hat{\mathbf{B}})=-\delta 
\Gamma ^{-}(\mathbf{p},\mathbf{q},\hat{\mathbf{B}})^{*} \),
so that \( \delta \Gamma ^{-}(\mathbf{p},\mathbf{q},\hat{\mathbf{B}}) \) must
be anti-hermitian. Mirror symmetry imposes in addition that \( \mathbf{T}_{\mathbf{pp}'}
(\mathbf{B})=\mathbf{T}_{-\mathbf{p}-\mathbf{p}'}(\mathbf{B}) \)
and thus \( \Gamma ^{-}(-\mathbf{p},-\mathbf{q},\mathbf{B})=\Gamma ^{-}(\mathbf{p},\mathbf{q}
,\mathbf{B}) \).
The general form of the tensor \( \delta \Gamma ^{+}(\mathbf{p},\mathbf{q},\hat{\mathbf{B}})
 \)
allowed by symmetry has already been discussed in Ref. \cite{euro} and will
be only slightly different for \( \delta \Gamma ^{-} \): 
\begin{equation}
\label{gamma}
\delta \Gamma _{ij}^{-}(\mathbf{p},\mathbf{q},\hat{\mathbf{B}})=ib_{1}\det (\mathbf{p},
\mathbf{q},\hat{\mathbf{B}})\delta _{ij}-i\varepsilon _{ijk}p_{k}\left[ ib_{2}(\hat{
\mathbf{B}}\cdot \mathbf{q})+ib_{3}(\hat{\mathbf{B}}\cdot \hat{\mathbf{p}})(\mathbf{p}\cdot
 \mathbf{q})\right] +ib_{4}\left[ p_{k}\Phi _{ki}q_{j}+q_{i}p_{k}\Phi _{kj}\right] ,
\end{equation}
 where the \( b_{i} \) are real-valued coefficients to be determined. 

The contribution of the first and the last scattering events is obtained by
multiplying the tensor for the ladder diagrams of Eq. (\ref{C}) by free Green
tensors \( \mathbf{G}_{0} \). This gives the incoherent contribution to the
coherent backscattering,

\begin{eqnarray*}
\mathbf{G}_{0}\mathbf{G}_{0}^{*}\cdot L^{+}(\mathbf{p},\mathbf{p}',\mathbf{q})\cdot 
\mathbf{G}_{0}\mathbf{G}_{0}^{*} & \simeq  & \mathbf{G}_{0}\cdot d(\mathbf{p},\mathbf{q})
\cdot \mathbf{G}^{*}_{0}\, \, \mathbf{G}^{*}_{0}\cdot d(-\mathbf{p}',-\mathbf{q})\cdot
 \mathbf{G}_{0},\\
 & \simeq  & (1-i\ell ^{*}\hat{\mathbf{p}}\cdot \mathbf{q})\widetilde{\rho }(\mathbf{q}=0
 ,z,z')(1+i\ell ^{*}\hat{\mathbf{p}}'\cdot \mathbf{q})\Delta _{\hat{\mathbf{p}}}\Delta
  _{\hat{\mathbf{p}}'},\\
 & \simeq  & \widetilde{\rho }\left( \mathbf{q}=0,z-\ell ^{*}\hat{p}_{z},z'+\ell ^{*}\hat{p}_{z}'\right) \Delta _{\hat{\mathbf{p}}}\Delta _{\hat{\mathbf{p}}'},
\end{eqnarray*}
in terms of the Fourier transform of the ladder propagator \( \widetilde{\rho }(\mathbf{q},z,z') \),
which is obtained from the solution of the diffusion equation (\ref{diffusion}).
The coherent contribution depends on the most-crossed diagrams \( C^{+}_{ijlk} \).
The tensor \( C^{+}_{ijlk} \) is obtained from \( L_{ijkl}^{-} \) after reversing
the indices \( k \) and \( l \), according to Eq. (\ref{CL}), and after adding
the contribution of the first and the last scattering event. In this calculation
of the tensor \( C^{+}_{ijlk} \), the reciprocity transformation (\ref{CL})
for the normal components \( z \) and \( z' \) of the tensor has been neglected.
The result is finally evaluated at reflection where \( z=z'=0 \). The components
of the ingoing and outgoing wave vectors \( \mathbf{p},\mathbf{p}' \) perpendicular
to the \( z \) axis are denoted by \( \mathbf{p}_{\bot },\mathbf{p}_{\bot }' \).
Finally the coherent contribution can be written 

\newpage

\begin{eqnarray}
C^{+}_{ijlk}(\mathbf{p},\mathbf{p}',\mathbf{p}+\mathbf{p}',\mathbf{B}) & = & \widetilde{\rho }
(\mathbf{p}_{\bot }+\mathbf{p}_{\bot }',-\ell ^{*}\hat{p}_{z},\ell ^{*}\hat{p}_{z}')\times 
\nonumber \\
\Delta _{il}\Delta _{jk} & + & i\left[ \Delta _{il}M^{-}_{jk}\left( \frac{\mathbf{p}_{\bot }'-
\mathbf{p}_{\bot }}{2},\mathbf{p}_{\bot }+\mathbf{p}_{\bot }',-\mathbf{B}\right) +M^{-}_{il}
\left( \frac{\mathbf{p}_{\bot }-\mathbf{p}_{\bot }'}{2},\mathbf{p}_{\bot }+\mathbf{p}_{\bot }',
\mathbf{B}\right) \Delta _{kj}\right] \label{dd} 
\end{eqnarray}
 with the definition
\begin{equation}
\label{M}
M_{ij}^{-}(\mathbf{p},\mathbf{q},\mathbf{B})=-\zeta \Phi _{ij}-\frac{\ell }{\omega }\Gamma ^{-}_{ij}(\mathbf{p},\mathbf{q},\mathbf{B}),
\end{equation}
 and \( \zeta =2(id_{2}-\eta ) \) is the Mie generalization of the parameter
\( F \) in Eq. (73) of Ref. \cite{bart}, which was shown to produce a rotation
of the polarization vector in the linear polarization channels of coherent backscattering.
To first order in the magnetic field, the front factor \( \widetilde{\rho }(\mathbf{p}_{\bot }+
\mathbf{p}_{\bot }',-\ell ^{*}\hat{p}_{z},\ell ^{*}\hat{p}_{z}') \)
of Eq. (\ref{dd}) is evaluated by replacing the parameter \( L_{a} \), which
was an absorption length for the evaluation of the light intensity in Eq. (\ref{diffusion}),
by a factor depending on the backscattering angle according to:
\[
\frac{1}{L_{a}^{2}}\rightarrow (\mathbf{p}_{\bot }+\mathbf{p}_{\bot }')^{2}.\]
 In second order in the magnetic field, \( \lambda  \) would be present here
as well as stated in Eq. (\ref{C}).

The values of the coefficients \( b_{i} \) can only be found by solving a system
of four coupled equations that one obtains when inserting Eqs. (\ref{dq},\ref{gamma})
into Eq. (\ref{BS}), and which is not reported explicitly here. The contribution
of \( b_{3} \) always vanishes in Eq. (\ref{dd}), since it is proportional
to the scalar product \( \mathbf{p}\cdot \mathbf{q} \) in Eq. (\ref{gamma}),
which is transformed into \( (\hat{\mathbf{p}}'-\hat{\mathbf{p}})\cdot (\hat{\mathbf{p}}+\hat{
\mathbf{p}}')=0 \)
in the operation involved in Eq. (\ref{CL}). For the same reason there is no
contribution of \( \Gamma ^{0}(\mathbf{p},\mathbf{q}) \) in Eq. (\ref{dd}).
Selection rules for the polarization are obtained in the helicity basis by considering
the product 
\begin{equation}
\label{rule}
C^{+}_{\sigma \sigma '}=\overline{P^{\sigma }_{ik}}(\hat{\mathbf{p}})C^{+}_{ijkl}P^{
\sigma '}_{jl}(\hat{\mathbf{p}}').
\end{equation}
 In this calculation, the terms proportional to \( \zeta  \) in Eq. (\ref{M})
disappear, as well as the contribution from \( b_{4} \) which is longitudinal
as can be seen from Eq. (\ref{gamma}). Among the four terms of \( \delta \Gamma (\mathbf{p},
\mathbf{q},\hat{\mathbf{B}})^{-}, \)
only the terms proportional to \( b_{1} \) and \( b_{2} \) survive, and Eq.
(\ref{fin}) is obtained for \( C^{+}_{\sigma \sigma '} \).

\newpage


\end{document}